\documentstyle[prl,aps]{revtex}

\begin{document}

\title{Textured Edges in Quantum Hall Systems}
\author{A. Karlhede$^1$, S. A. Kivelson$^2$, K. Lejnell$^1$ and 
S. L. Sondhi$^3$}

\address{
$^1$Department of Physics,
Stockholm University,
Box 6730, S-11385 Stockholm,
Sweden
}

\address{
$^2$Department of Physics,
University of California Los Angeles,
Los Angeles, CA 90024, USA
}

\address{
$^3$Department of Physics,
Princeton University,
Princeton, NJ 08544, USA
}

\date{\today}
\maketitle
\begin{abstract}

We have investigated the formation of spin textures at the edges of
quantum Hall systems for several ferromagnetic filling factors.
The textures are driven by the same physics that leads to ``skyrmions''
in the bulk.
For hard confinement and large Zeeman energies, the edges are narrow and
spin polarized. Away from this limit and for $\nu=1/3$, $1/5$ and $1$, we
find that they widen by developing spin textures. In contrast, for $\nu=3$,
the edge remains spin polarized, even as it reconstructs.
We comment on the mode structure of the reconstructed edges and
on possible experimental signatures.

\end{abstract}

Progress in device fabrication and theoretical interest in moving
beyond the basic physics of the quantum Hall effect (QHE), have
recently converged in a flurry of activity on multicomponent systems
\cite{gmreview}. 
One of the new themes in this work is the existence of excitations
that involve topologically nontrivial configurations of the 
components. The first examples to be uncovered were ``skyrmions''
of the spin degrees of freedom \cite{sondhi1} and there is now
a considerable amount of experimental evidence for their existence
\cite{barrett1,schmeller,goldberg}. 
Meanwhile, the Indiana group has studied pseudospin textures in double 
layer systems \cite{indiana} and used them to
account for a novel phase transition observed as a function of a
parallel magnetic field \cite{murphy}.

The physics here is a combination of 
magnetism, Berry phases, and the density-flux commensuration that
is at the heart of the QHE. As a consequence,
at ferromagnetic QH fillings, long wavelength variations in the 
density are most cheaply accomplished by texturing
the spins \cite{sondhi1,indiana}. In this letter we discuss another instance
where this tradeoff might be expected to exist, i.e. the edges
of QH systems.
For a confining potential that rises steeply at the edge, the density
falls steeply from its bulk value and the width of the edge region
is of order the magnetic length $\ell=\sqrt{\hbar c/eB}$; 
most work on edge states 
\cite{wenreview} assumes this simple structure. As the confinement
is softened, the edge will reconstruct by the outward motion of
the charge and will become wider in the process 
\cite{macdonald1,chamon,reconstruct}.
The precise question we address is the following: for the edges of 
ferromagnetic QH states \cite{ferroedges}, under what circumstances
does edge reconstruction take place by texture formation?

A subtlety here \cite{xgwu}, is that the physics of textures is most 
evident
for long wavelength density fluctuations. Whether it
survives at wavelengths of the order of $\ell$ is an issue of
detail and must be settled by explicit calculations.
Consequently, we have searched for edge texture formation using both 
Hartree-Fock and effective action based calculations at $\nu=1$, $3$, $1/3$
and $1/5$. In addition to the width of the edge region, the other important
parameter in the problem is the ratio, $\tilde g
 \equiv g \mu_B B/ (e^2/\epsilon \ell)$, of the Zeeman energy,
$g \mu_B B$, to
the typical Coulomb energy, $e^2/\epsilon \ell$;
clearly, for large enough $\tilde g$ all states are spin polarized.

Our chief results are the following. 
{\it (i)} At $\nu=1$ and for $\tilde g$ less 
than a critical value $\tilde g_c$, the formation of an edge texture 
preempts the 
polarized instability\cite{macdonald1,chamon}. 
{\it (ii)} In contrast, at $\nu=3$, the edge remains polarized even 
as it reconstructs.
This distinction between $\nu=1$ and $\nu=3$
parallels that for skyrmions in the bulk \cite{xgwu} and
strongly suggests that spin textures govern the low energy
physics at small Zeeman energies
only for states in the Lowest Landau level,  {\it i.e.}
for $\nu=1$, $\nu=1/3$, and $1/5$.
{\it (iii)} The polarized edge is invariant under
spin rotations about the magnetic field as well as under
translations along the edge. The textured edge is invariant only under a
specific linear combination of these symmetries whereas the orthogonal
combination is spontaneously broken. As a consequence, we expect that
the reconstructed edge has an additional gapless mode.
We now turn to the details of our results.

\noindent
{\bf Edge Hamiltonian and Polarized Instability:}
We consider a two dimensional electron gas in a magnetic field ${\bf B}$
and restrict the orbital Hilbert space to the highest occupied Landau level,
$n$. 
The Hamiltonian is constructed by taking matrix elements of the Coulomb
interaction $V({\bf r})=e^2/(\epsilon |{\bf r}|)$ between states in the
$n^{th}$ Landau level, in the presence of a specified background charge
$\rho_b(\bf r)$ which confines the electron gas.
(For $n>0$, $\rho_b$ is taken to include the charge
in all lower, fully occupied Landau levels.) 
We focus, in the present paper, on a 
semi-infinite system with a straight edge; similar effects
in quantum dots will be discussed elsewhere \cite{kennet}. 
For this geometry, it is convenient to work in Landau gauge 
(${\bf A} = Bx {\bf \hat{y}}$) with periodic boundary conditions
in the ${\bf \hat{y}}$ direction;
the field operators can thus be expanded as
$\psi_{\sigma}=\sum_p \varphi_p c_{p\sigma}$, 
where $c_{p \sigma}$ 
annihilates an electron with wavevector $p$ and spin $\sigma$. For the $n+1$st 
Landau level and a system of finite length $L$ along the edge,
$\varphi_p = (2^n n! \sqrt{\pi} L \ell)^{-1/2} e^{ipy} 
H_n(\frac{x}{\ell} - p\ell)
e^{-(\frac{x}{\ell} - p\ell)^2/2}$, where $H_n$ is the $n$th Hermite polynomial
and $p=2\pi n_p/L, \, n_p=0, \pm 1, \pm 2$, etc. We take 
$\rho_b$ to fall 
linearly from its bulk value $\bar \rho$ to zero over a region
of width $w$ (Fig 1), so the  
``softness'' of the edge is set by the dimensionless ratio $\tilde{w}
\equiv w/l$.

Chamon and Wen 
studied the stability of the  straight $\nu = 1$ edge as a function of $\tilde w$ 
for polarized electrons \cite{chamon}. For small 
$\tilde w$ the edge is sharp; the occupation of the up spin one-particle
 states is 
unity out to a maximum wavevector and zero thereafter hence the 
density falls to zero over a distance  of $O(\ell)$ in real space. 
At $\tilde{w} = \tilde{w}_{p} = 9.0$ (in Hartree-Fock) 
an  ``edge reconstruction'' occurs where
a lump of QH liquid is split off and deposited a distance $\sim \ell$ from 
the bulk.  

\noindent  
{\bf Edge Textures:} 
We study the stability of polarized edges against formation of spin textures 
by two different methods. The first utilizes an effective action for the 
spin degrees of freedom at ferromagnetic QH states obtained after the charge 
dynamics have been integrated out. The Lagrangian
density has the form\cite{sondhi1,indiana},
\begin{eqnarray}
& &{\cal L}_{eff} =  \frac 1 2 \bar\rho  A({\bf n}) \partial_t {\bf n} -
        \frac 1 2 \rho^s ({\bf \nabla n})^2
        + g \bar\rho \mu_B {\bf n B}  -  \frac{\nu_{\rm FM}^2}{2} \\
& &  \times
\int d^2 r' \,[\bar\rho-q({\bf r})-\rho_b({\bf r})] V({\bf r} - {\bf r'}) 
[\bar\rho-q({\bf r'})-\rho_b({\bf r'})]    \ , \nonumber
\label{seff}
\end{eqnarray}
where the electron density is the difference $\bar\rho-q$,
$\rho^s$ is the spin stiffness \cite{rho_s} and $\nu_{\rm FM}$
is the filling factor of the ferromagnetic component
({\it e.g.} $\nu_{\rm FM}=1$
at $\nu=3$). 
The spin is described by the
unit vector ${\bf n}({\bf r})$, ${\bf A}({\bf n})$ is
the vector potential of a unit monopole, i.e., 
$\epsilon^{abc}\partial_b A_c = n^a$ and $q({\bf r}) = 
{\bf n} \cdot (\partial_x {\bf n} \times \partial_y {\bf n})/
4 \pi$ is the topological (Pontryagin) density of the spin field.

The applicability of this effective action to the edges of a quantum
Hall system is not evident. The derivation presented in \cite{sondhi1}
assumed gapped density fluctuations 
and the absence of currents, both of 
which are problematic at the edge. It is
our belief, supported by the results of microscopic
calculations presented later, that for the {\em statics} of the
textured edge this is not a serious limitation. The 
energetics of the textures are quite robust and are captured well
by the static terms in the effective action. 

Edge textures are configurations of the spin field 
that possess a topological density at the edge of the system. Denoting
the directions perpendicular and parallel to the edge by $x$ and $y$ 
and the magnetic field axis by $z$, the spin field takes the form,
\begin{eqnarray} \label{ansatz}
n_x & = & \sqrt{1-f^2(x)} \ {\rm cos}(ky+\theta_0)  \ , \ \\
n_y & = & \sqrt{1-f^2(x)} \ {\rm sin}(ky+\theta_0)  \ , \ \ \
n_z = f(x) \ \ \  ,\nonumber 
\end{eqnarray}
where $k$ and $\theta_0$ are constants and $f(x)$ falls from its value
$f(-\infty)=1$ in the bulk polarized state as the edge is approached.
Evidently, the spins tilt away from the direction of the field
on going across the edge while they precess about it with wavevector
$k$ along the edge; the trivial case, $f \equiv 1$, is the polarized edge
(Fig 1).
The topological density of the texture,
$q=-(k/4\pi) df/dx$, is proportional both to the gradient of 
$n_z$ and to the edge wavevector. 
Note that although $q$, and hence the electron
density $\rho=\bar \rho-q$, is constant along the edge, the form
(\ref{ansatz}) breaks
translational invariance along the edge, generated by $t_y=-i\partial_y$,
as well as spin rotational symmetry about the magnetic field generated by
$L_z = \sigma_y$ acting on $(n_x, \, n_y)$. 
However, the state (\ref{ansatz}) is invariant under the
combined symmetry generated by $t_y+kL_z$. Thus there is one broken
continuous symmetry and the symmetry related states are labelled by $\theta_0$.

To find the optimal texture, we  determine 
$f(x)$ by numerically integrating the equation of motions
using a relaxational procedure for a fixed $k$, which is then fixed by
minimizing the energy.

\noindent
{\bf Hartree-Fock:} The second method used to search for textured edges
is the Hartree-Fock technique introduced by Fertig 
{\em et al.} for skyrmions \cite{fertig,karlhede}. The form of the wavefunction
is fixed, by the constraints of lowest Landau level occupation and the
conservation of $t_y+kL_z$ noted earlier, to be
\begin{equation}
|k\rangle = \prod_p (u_p c^{\dagger}_{p\uparrow} + 
v_p c^{\dagger}_{p+k\downarrow}) |0 \rangle	\ \ \  .
\label{HFansatz}
\end{equation} 
The coefficients $(u_p, v_p)$ are determined by numerically iterating the 
Hartree-Fock equations until a self-consistent solution is found.
The wave vector $k$ is again determined by minimizing 
the energy. Recovering the polarized bulk state requires that 
$u_p \rightarrow 1$ deep in the bulk but its values at the edge are
unconstrained. The numerical results in this paper are for 
a system of length $L=100 \ell$; we have checked that they are not sensitive
to this choice.

\noindent
{\bf Results:} We begin with our results on
$\nu=1$ which are illustrated in Figs. \ref{phase} \& \ref{profiles}.
Within one Landau level the dimensionless parameters $\tilde g$ and
$\tilde w$ control the physics. 
Fig. \ref{phase} shows the stability diagram of the edge in the 
$(\tilde g, \tilde w)$ plane, with phase boundaries obtained both from 
Hartree-Fock and effective action calculations. The close agreement
between these methods at small $\tilde{g}$ where the textures are expected
to be slowly varying is encouraging and in line with similar calculations 
for bulk skyrmions \cite{kennet}. The Hartree-Fock treatment
represents the sharp edge exactly and hence gives an upper bound for the
critical $\tilde{w}$ at a fixed $\tilde{g}$.

Notice that for small Zeeman energies,
$\tilde g <\tilde  g_c= 0.082$, and softening confinement, the sharp edge becomes unstable 
to a textured reconstruction {\em before} it becomes unstable to 
polarized reconstruction. In particular, at $\tilde g =0$, the transition to a 
textured edge with wavevector $k=0.63$ occurs at $\tilde w=6.7$ while the
purely polarized reconstruction only happens at $\tilde w_p=9.0$.
As $\tilde g$ increases, the critical $\tilde w$ and $k$ increase as
well till finally at $\tilde g = \tilde  g_c$, the polarized instability becomes
the primary instability. The quoted numbers are from Hartree-Fock; the 
effective theory gives that the $\tilde g = 0$ transition occurs at 
$\tilde w = 6.9$ with $k=0.40$. We remind the reader that the effective theory
does not allow a computation of the polarized instability and hence of $\tilde  g_c$.
In contrast the Hartree-Fock wavefunctions allow for both possibilities.

The reconstruction in going across the phase boundary is continuous, 
as illustrated in Fig.~\ref{profiles} where we show (two-dimensional) density
and spin-density profiles. The density profile evolves
smoothly from that of the sharp edge towards the fragmented edge produced by 
polarized reconstruction. As the initial instability involves moving a small
amount of charge locally, textured reconstruction is really an edge instability;
in contrast, for $\tilde g > g_c$,
the polarized instability sets in when the edge is still locally
stable \cite{macdonald1,chamon}. The inset table in
Fig.~\ref{profiles} details the
increase in the edge wavevector and the depolarization of the 
edge away from the phase boundary.

For $\nu \neq 1$, it is straightforward to extend the effective action calculations 
and one again finds instabilities of the sharp edge
to texturing (Fig. 2). At $\tilde g=0$ these set in at critical values 
$\tilde w _{\nu}$ where 
$\tilde w_{1/3}=2.4$,  $\tilde w_{1/5}=1.7$ and 
$\tilde w_{3}=12$. The corresponding edge wavevectors are $1.11,\,  1.57$ 
and $0.23$. 
One notes that 
$\tilde w_\nu$  are related to the value at $\nu=1$ by 
$\tilde w_\nu \approx (1/\nu_{\rm FM}^2)(\rho^s_\nu/\rho^s_1)\tilde w_1$, where
$\rho^s_\nu$ is the spin stiffness at filling factor $\nu$.

For $\nu=3$, we
have carried out an analogous calculation to that of Chamon and Wen 
to find the critical value of $\tilde w_p$ at which the polarized
instability occurs.  We find that $\tilde w_p = 8.3$ which
is obviously less than $\tilde w_{3}$.  (This is corroborated
by the fact that no texturing is observed in the Hartree-Fock
solutions for $\tilde w < \tilde w_p$, even at 
$\tilde g=0$.)
Consequently, the $\nu=3$ edge (really,
the innermost of its three edges) will not display texturing and should reconstruct 
to a polarized, fragmented edge with softening confinement.

For $\nu=1/3$ and $1/5$ no estimate for a polarized instability is 
available and we do not know if it preempts texture formation.
Nevertheless, the moral of the $\nu=3$
calculations is that the relative energetics of the polarized
and textured instabilities roughly follow those of polarized quasiparticles
and skyrmions \cite{xgwu}. This strongly suggests that they exhibit textured edges
at small $\tilde g$ as well.

\noindent
{\bf Edge Dynamics:} 
The limitations of the effective action, detailed earlier, preclude
a proper treatment of the dynamics at this time; hence,
we limit ourselves to a qualitative discussion.

Close to the phase boundary, the most striking attribute
of the mean field solutions is the presence of a broken
symmetry. As this is a broken continuous symmetry in a one dimensional 
quantum system,
it is likely that the true long range order in these solutions will be reduced
to quasi (algebraic) long range order when quantum fluctuations are included. 
Consequently,
the true ground state will conserve both $t_y$ and $L_z$ although the 
transverse
spin-spin correlation function will show evidence of the textures 
discussed in this paper.
A true broken symmetry would imply an extra gapless, Goldstone mode and 
we expect that
a modified version of this will survive fluctuations as well.

\noindent
{\bf Experimental Consequences:} At $\nu=1$, the range of Zeeman energies for which there is 
an instability to textured reconstruction is quite large, indeed it covers all
values of interest for GaAs systems ($0.005 < \tilde g < 0.02$ for fields
between $1T$ and $10T$). The major experimental barrier to
realizing the textured edge is tuning the confining electric field to the right 
range. Recent theory \cite{multichannels} suggests that standard samples may contain highly
reconstructed edges with several channels even for integer states. If this is
correct then it might be necessary to investigate systems with ultra-sharp
confinement produced by cleaving which have been fabricated recently \cite{cleave}.

The principal signatures of textured edges are likely to be
their sensitivity to the value of the Zeeman energy and the associated 
depolarization
of the edge. The former could be investigated by tunneling into the edge of the
electron gas 
at various values of a tilted magnetic field.
The latter might be amenable to a local NMR probe.
As
with skyrmions, the contrast between $\nu=1$ and $3$ 
should prove to be a useful diagnostic. 

Finally, we note that spin effects in the reconstruction of small
droplets were found in exact diagonalization studies 
of small systems by Yang, MacDonald 
and Johnson \cite{myj}. 
As we were writing up this work we
received a preprint by Oaknin {\em et al.} \cite{oaknin} which discusses the 
existence of a branch of textured excitations for QH droplets 
with $\nu \simeq 1$.

\acknowledgements
We are grateful to L. Brey, H. A. Fertig, S. M. Girvin, T. H. Hansson, A. H. MacDonald and
P. L. McEuen for useful discussions. This work was supported in part by NSF 
grant \# DMR 93-12606 (SAK) and the Swedish Natural Science Research Council 
(AK).

\begin{figure}
\caption{
Sketch of $\rho_b$ and $n_z$ perpendicular to the edge. The inset shows how 
the planar components of the spin rotate along the textured edge.
}
\label{edge}
\end{figure}

\begin{figure}
\caption{
Stability diagram for $\nu=1$ edge showing the region of stability of the
sharp edge and the regions where it is unstable to texture formation
and to a polarized reconstruction. The solid line
marks the onset of the textured edge within Hartree-Fock and the dashed
line within the effective action calculations. The sharp edge becomes
unstable to a polarized reconstruction for $\tilde{w} > \tilde{w}_p 
= 9.0$ (dotted line) independent of $\tilde{g}$. The wavevectors for
the initial textures are indicated along the boundary. Also included are  
effective action predictions for the onset of the textured edge at 
$\nu=1/3$ and $1/5$ (dashed lines).
}
\label{phase}
\end{figure}

\begin{figure}
\caption{
Density and spin density as a function of $x$ (distance perpendicular
to the edge) for $\nu=1$, $\tilde g = 0.01$ and a set of
$\tilde w$. The curves labeled $a$ are for the polarized edge while $c$ 
are for a textured edge beyond the polarized instability where the
former is still energetically preferred.
The table shows the wave vector $k$, as well as the 
spin, $S/L$, and energy, $E/(Le^2/\epsilon \ell)$, per unit length
relative to the polarized edge.
}
\label{profiles}
\end{figure}


\begin{references}
\bibitem{ferroedges} The same question arises for non-ferromagnetic states,
e.g. $\nu=2$, which are hierarchical descendents of ferromagnetic states 
and hence contain them in their edge structure \cite{kennet}.
\end{references}
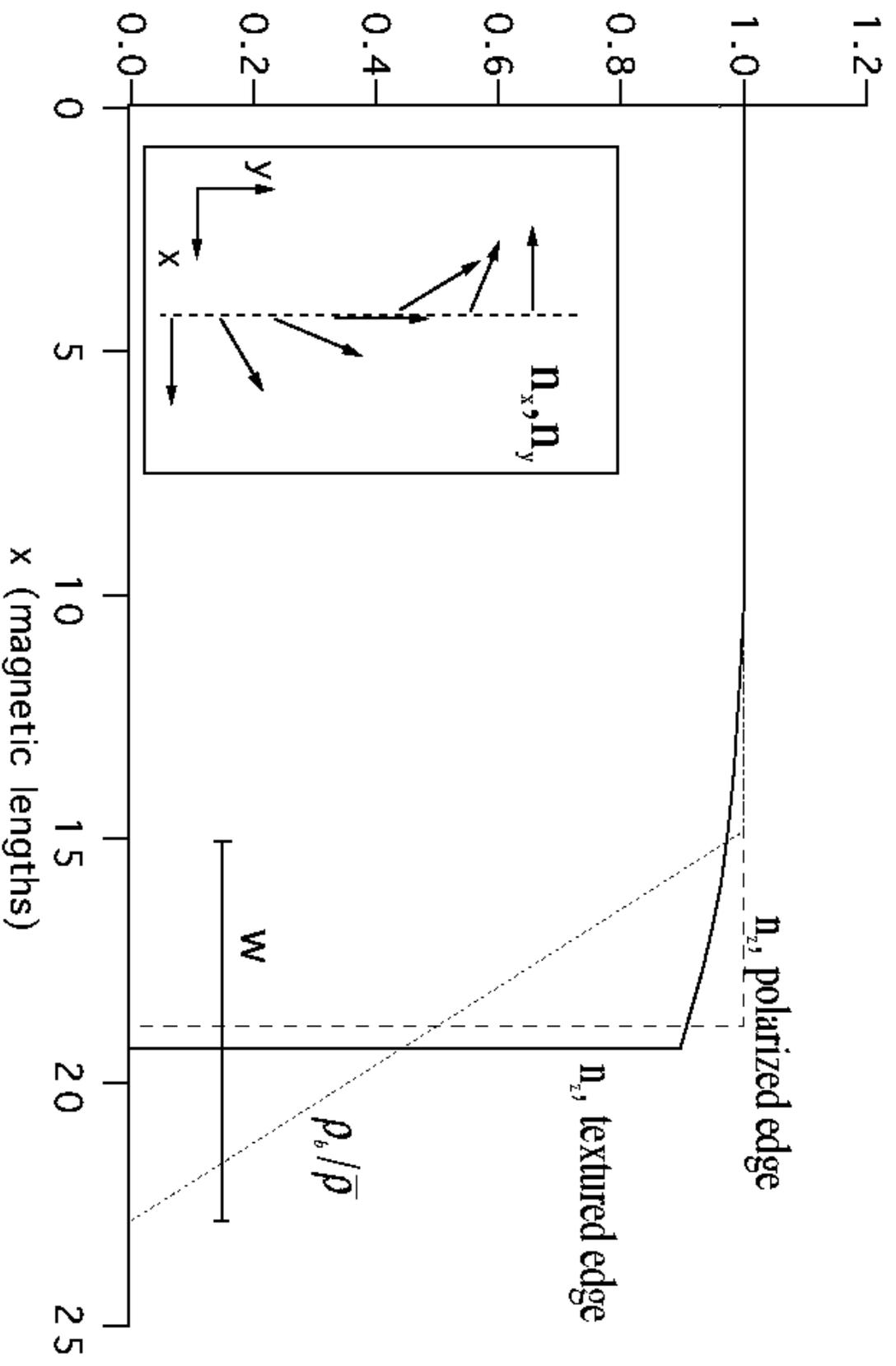
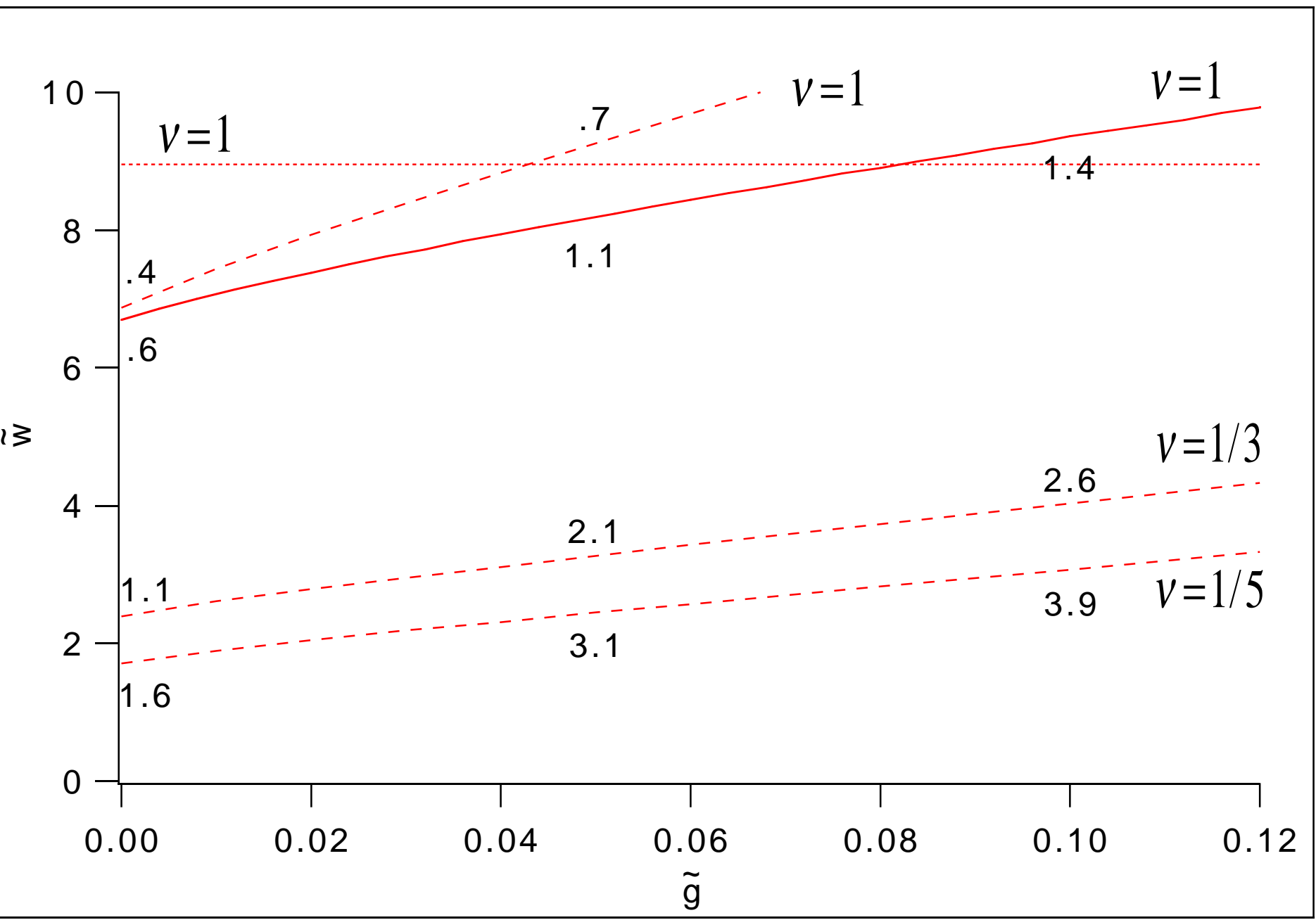
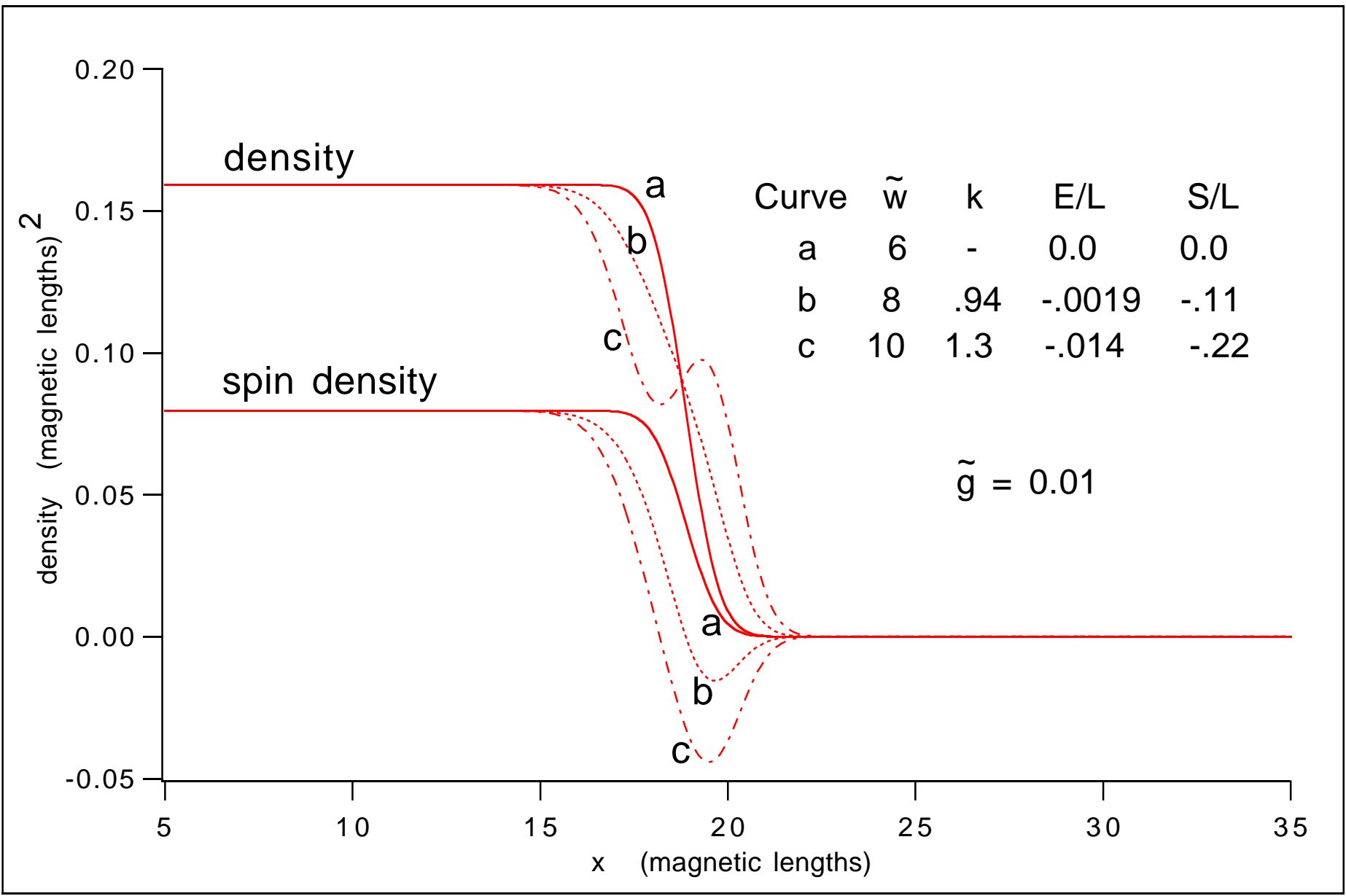
\end{document}